# Insights into the Second Law of Thermodynamics from Anisotropic Gas-Surface Interactions

S. L. Miller [a]

Sandia National Laboratories [b]
Albuquerque, NM  87185

Thermodynamic implications of anisotropic gas-surface interactions in a closed molecular flow cavity are examined. Anisotropy at the microscopic scale, such as might be caused by reduced-dimensionality surfaces, is shown to lead to reversibility at the macroscopic scale. The possibility of a self-sustaining nonequilibrium stationary state induced by surface anisotropy is demonstrated that simultaneously satisfies flux balance, conservation of momentum, and conservation of energy. Conversely, it is also shown that the second law of thermodynamics prohibits anisotropic gas-surface interactions in "equilibrium", even for reduced dimensionality surfaces. This is particularly startling because reduced dimensionality surfaces are known to exhibit a plethora of anisotropic properties. That gas-surface interactions would be excluded from these anisotropic properties is completely counterintuitive from a causality perspective. These results provide intriguing insights into the second law of thermodynamics and its relation to gas-surface interaction physics.

## I. INTRODUCTION

The fundamental nature of gas-surface interactions remains an important, but often elusive, area of research, with new discoveries and model refinements continually being made [1-8]. Much of what we do understand regarding gas-surface interactions is founded on the paradigm of a chaotic gas in contact with a bounding surface, such as in a viscous flow cavity where randomizing intermolecular collisions dominate the gas dynamics. The inferred gas-surface interaction mechanisms are typically constrained to satisfy boundary conditions that necessarily preserve the essential properties of the chaotic gas ensemble. Thus, the properties of the gas ensemble itself play a dominant role in guiding our understanding of the associated gas-surface interaction physics.

For a molecular flow cavity, where the mean free path is much longer than a characteristic dimension of the cavity, the situation must be reversed. The chaotic hypothesis is not a universally justifiable a-priori assumption [9,10]. This is because the properties of the gas ensemble are determined by the nature of gas-surface interactions. For example, isotropy and homogeneity, when they occur in molecular flow systems, cannot directly result from a preponderance of randomizing collisions between gas molecules, but must result from other principles related to gas-surface interactions. That has historically not been an issue because surfaces have typically been presumed to be inherently randomizing. It is now realized that assumption may not be universally valid, considering the large number of anisotropic properties that have recently been discovered associated with nano-engineered reduced-dimensionality surfaces [11-16]. While such surfaces may reasonably be anticipated to impact an associated molecular flow gas distribution, the second law of thermodynamics requires irreversibility, and by inference, homogeneity and randomness. Thus, the molecular flow cavity forms an intriguing paradigm where the second law of thermodynamics and gas-surface interaction physics are inextricably linked in a unique and profound way. It is this observation that motivates this paper.

[a]  e-mail: sam.miller@comcast.net
[b]  Employer.  Sandia National Laboratories is not officially affiliated with this work.

The purpose of this paper is to examine properties of gas cavities dominated by surface interactions, and to understand those properties in relation to the second law of thermodynamics. In particular, we wish to explore the question "must all surfaces, including reduced-dimensionality surfaces, be inherently randomizing?" To do this, we geometrically relate the incoming molecular distribution impinging on a surface element to the outgoing distributions leaving surface elements elsewhere on the surface of a molecular flow cavity. This is fundamentally different from the more conventional scattering kernel-based approach relating an outgoing distribution to an incoming distribution at the same surface element location [1,3,5,17]. Consequences of flux balance, conservation of momentum, and conservation of energy are investigated. Inevitably, the discussion leads to the cosine distribution of molecular flux, and the exposure of errant perspectives regarding the interpretation of previous derivations of the cosine distribution. Two fundamental, but mutually exclusive, outcomes present themselves. These outcomes comprise two foundational opportunities for further contribution by the motivated reader.

This paper is structurally organized with mathematical details confined to Appendices, permitting the reader to more easily focus on the conceptual principles being discussed.

## II. SYSTEM DEFINITION/BACKGROUND

To clearly define the system under consideration, it is worthwhile to first carefully address the question of what is the fundamental role of a surface in relation to gas-surface interactions. To do this, we consider a surface element during the process of a gas cavity transitioning from one state of equilibrium to another state of equilibrium. We refer to the cavity of Fig. 1, which is bounded by a thermally conductive wall having an inner surface and an outer surface. The temperature of the outer surface may be externally controlled. In an initial state of equilibrium, a membrane confines gas molecules to the volume indicated. The membrane is then breached, allowing molecules to freely redistribute throughout the entire cavity, where a second state of equilibrium is reached after a period of time.

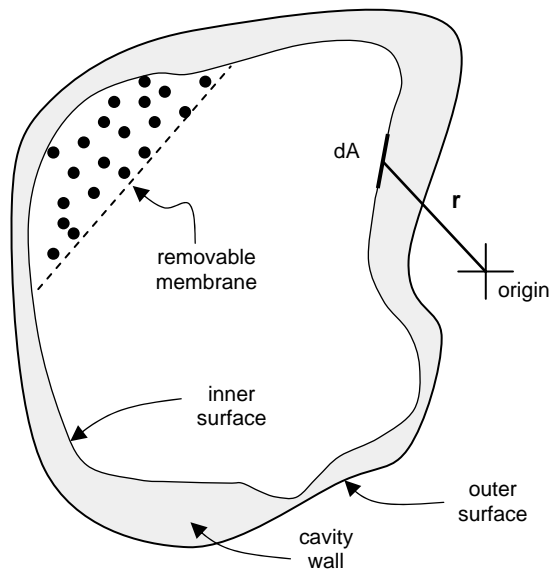

**FIG. 1.** An arbitrarily shaped cavity in the molecular flow regime. Surface element *dA* receives and redirects molecular flux distributions when the membrane is removed.

For any short time interval during the nonequilibrium transition period, surface element *dA* at location *r* converts an incoming molecular distribution to an outgoing distribution. In general, the outgoing angular flux distribution is different from the incoming angular flux distribution, and both change with time. In addition, energy and momentum may be exchanged between gas molecules and the surface element. Equilibrium principles such as detailed balance and reciprocity [1,17-19] are not satisfied (and are not applicable) during this nonequilibrium transition period.

The properties of the outgoing and incoming molecular distributions (e.g. angular flux, energy, and momentum distributions) are related by the physics of gas-surface interactions and boundary conditions for those interactions. The surface is considered to be anything that redirects the trajectory of incoming molecules, and hence may also include adsorbed species. The interactions of individual molecules with the surface may span the continuum from short duration interactions with no energy exchange (e.g. specular reflection) to long duration interactions with energy exchange (e.g. adsorption followed by desorption). The interactions may affect the effective surface coverage of adsorbed species. Boundary conditions for the interactions change during the transition between the two states of equilibrium.

It is important to note that surface element *dA* does not know the state of the entire system, e.g. it doesn't know whether or not the system is in equilibrium. The physics of the gas-surface interactions at *dA* does not change as the state of the system changes – only the boundary conditions for the interactions change. When the boundary conditions cease changing with time, that condition is a stationary state. If the boundary conditions of a stationary state satisfy various criteria related to the second law of thermodynamics (e.g. no energy transfer, detailed balance satisfied, etc.), then the surface element is said to be in equilibrium with the gas.

If the cavity were in the viscous-flow regime, the properties of the incoming molecular distribution (comprising a portion of the boundary conditions for the subsequent gas-surface interactions) would result directly from randomizing collisions between gas molecules. Specifically, the randomizing collisions would directly result in an incoming molecular distribution exhibiting isotropy, homogeneity, and the Maxwell speed distribution.

In contrast, in the molecular-flow regime the properties of the incoming molecular distribution striking *dA* are determined by the outgoing molecular distributions originating from all line-of-sight surface elements elsewhere on the cavity surface, and the distance and orientation of those surface elements. Consequently, determining the behavior of gas in a molecular flow cavity becomes a highly convoluted problem. Outgoing distributions depend on the physics of gas-surface interactions and the properties of incoming distributions. The incoming distributions are themselves outgoing distributions convolved with the shape of the cavity. Making quantitative statements regarding molecular flow distributions would seem to be an impossibly complex problem. Fortunately, it is not.

There are several approaches that have historically been taken to determine the physically permissible configurations of a molecular flow cavity. Unfortunately, these approaches tend to be very constrained by a-priori assumptions applicable to the traditional paradigm of a chaotic gas or viscous flow cavity. One approach involves using scattering kernels that relate properties of the outgoing gas distribution at a surface element to properties of the incoming distribution [1,3,5,17], where properties of the scattering kernel are typically constrained by principles such as detailed balance or reciprocity (DB&R) [1,17-19], or by the assumption of a Maxwellian distribution [20]. The applicability of DB&R or the Maxwell distribution explicitly depends on certain other conditions being met – conditions that we will not a-priori assume for the present analysis. Another approach is to use the Boltzmann transport equation as a starting point for determining gas configurations [5,6,21,22]. Unfortunately, the Boltzmann equation itself is derived using assumptions that immediately constrain the form of the molecular distributions. For example, the Boltzmann equation is derived assuming that the gas is chaotic, and that interactions with a surface preserve this assumed property. Because gas distributions in a molecular flow cavity are governed by gas-surface interactions, a-priori assumptions regarding the form of the gas distributions will be avoided in the present analysis. Instead, a simple geometric analysis method is used in conjunction with conservation principles.

We proceed by developing certain properties of a conventional equilibrium system, which also serves to introduce essential definitions and terminology. Next, we relax the requirement for the system to be in equilibrium, and examine more general stationary state systems. An insightful example of a nonequilibrium stationary state solution is given for a hemispherical cavity. The subsequent discussion focuses on what fundamental principles constrain the form of the outgoing angular flux distribution. Finally, an argument is given supporting the possibility of a self-sustaining nonequilibrium stationary state system.

## III. EQUILIBRIUM STATIONARY STATES

We begin with the special case where we do, in fact, make assumptions regarding the molecular distribution. For this special case illustration, we assume that the molecular speed distribution is isotropic and homogeneous. Specifically, within a given small spherical volume element within the cavity, we assume the average number of molecules traveling at a certain speed in one direction is the same as that traveling in any direction, and is the same independent of the location of the differential volume.

Molecular speed isotropy in a molecular flow cavity implies that two surface elements *dA* and *dA'*, as shown in Fig. 2(a), must exchange the same molecular flux (here we define flux simply as the number of molecules per unit time). This is because the number of molecules that flow in all possible paths from *dA* to *dA'* is balanced by the same number of molecules flowing in the opposite path directions. Homogeneity requires this balance of flux to occur between all possible pairs of surface elements.

These flux balance conditions uniquely constrain the form of the angular flux distribution for the entire cavity. To see this, we first introduce an outgoing angular flux distribution function $P(r,\theta_r,\phi_r)$ that represents the number of molecules leaving a surface at location *r* per unit time per unit area per unit solid angle in the direction of $\theta_r$ and $\phi_r$. As illustrated in Fig. 2(b), the angles $\theta_r$ and $\phi_r$ are referenced to a local coordinate system at the location *r* of the surface element – hence the subscripts *r*. As shown in Appendix A, both the outgoing and incoming angular flux distribution for each and every surface element of the cavity, as a result of the above flux balance conditions, must be

$$P(r, \theta_r, \phi_r) = K \cos(\theta_r), \tag{1}$$

where *K* is a constant for the cavity. Note that the angular flux distribution is independent of location on the cavity surface, and has no azimuthal dependence. Equation (1) is solely a geometrical result applicable to any molecular flow cavity where the speed distribution of the gas is assumed to be isotropic and homogeneous. Equation (1) is recognized to be the same as the familiar, but empirical, Knudsen cosine law of diffuse scattering [23].

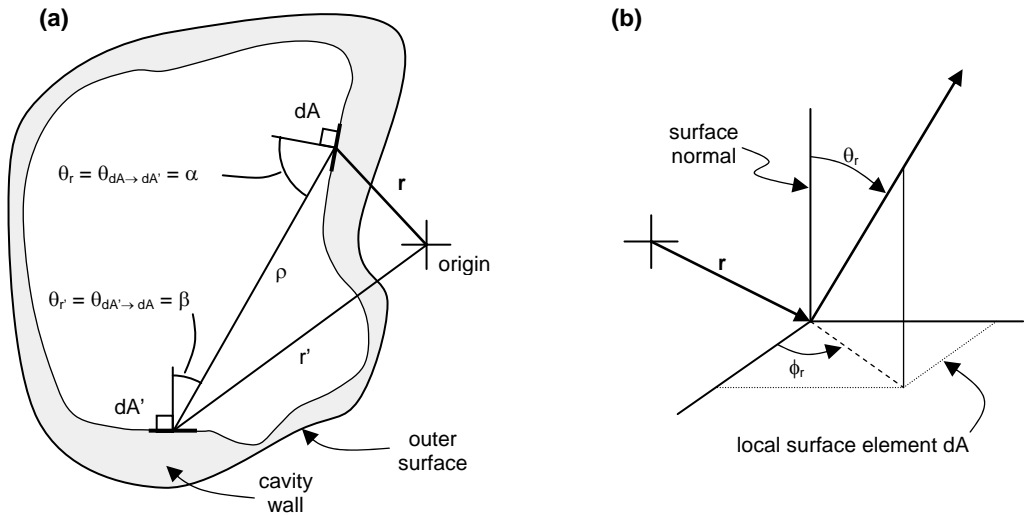

**FIG. 2.** (a) Geometric definitions used for flux balance calculations in an arbitrarily shaped cavity. (b) In the local coordinate system at each surface element, the angle $\theta$ is measured relative to the surface normal.

The physical cause and effect (causality) relationship is worth noting, and can involve two perspectives:
a) The second law of thermodynamics requires the gas to be isotropic and homogeneous, and hence requires the outgoing angular flux distribution to be cosine.
b) Isotropy and homogeneity of the speed distribution, when they occur in a molecular flow cavity, do not result from randomizing collisions between gas molecules. Instead, they must result from the existence of an outgoing angular flux distribution that is, for some reason, uniformly proportional to cos($\theta$) over the entire cavity surface. That reason must be directly related to the gas-surface interaction physics.

The following sections aim to provide clarifying insight into these two very distinct perspectives.

## IV. NONEQUILIBRIUM STATIONARY STATES

### A. General results

We now drop the step of a-priori specifying a property of the gas distribution, and see what may reasonably be concluded using the same geometric approach. Rather than requiring flux balance between any pair of surface elements (an assumed property of the gas distribution), we now instead only require flux balance at each surface element. By the phrase 'flux balance at a surface element' we simply mean that the incoming flux of molecules striking the surface element is the same as the outgoing flux of molecules leaving that same element. (Here we are explicitly assuming no net lateral transport of gas molecules that are adsorbed on the surface. This simplifies the analysis, but does not fundamentally affect the results and conclusions of the paper.) No assumption is made regarding angular flux distributions, or how flux densities (i.e. flux per unit area) may or may not change with location on a surface. We now consider how this sole assumption of flux balance at each surface element constrains outgoing angular flux distributions and flux densities for stationary states.

The first key result, shown in Appendix B, is that flux balance can always be satisfied at each surface element without requiring the flux density to be the same at all locations, and without requiring any particular form for the outgoing angular flux distribution. Specifically, any outgoing angular flux distribution, where the form of the angular dependence varies with location, can be made to satisfy flux balance for all locations of a cavity surface simply by including an overall scaling factor that varies with location. This scaling factor is unique for any given location-dependent outgoing angular flux distribution. Expressed mathematically, given some function $x(\bm{r},\theta_r,\phi_r)$, there exists a unique corresponding scaling function $y(\bm{r})$ such that the outgoing angular flux distribution function $P(\bm{r},\theta_r,\phi_r) \equiv y(\bm{r})x(\bm{r},\theta_r,\phi_r)$ satisfies flux balance at every surface location $\bm{r}$. Physically, this means that the flux density (and hence pressure) may vary with location for such a solution.

It is essential that such mathematical solutions exist, because nonequilibrium stationary state systems do indeed physically exist. A nonequilibrium stationary state may be achieved for the cavity in Fig. 2(a) as follows. When a thermal gradient is externally maintained along the outer surface of the cavity wall, the system comprising the gas and inner cavity surface will reach a stationary state, with a thermal gradient also existing along the inner cavity surface. As the above results show, a nonuniform flux density could very well (and perhaps must) accompany such a stationary state, while satisfying flux balance at every surface element on the cavity. The position-dependence of the flux density will depend on the geometry of the cavity, as well as other quantities.

A nonequilibrium stationary state is now further examined in the context of conservation of momentum and energy. Using the same approach as used in Appendix B, it can be shown that any set of momentum boundary conditions may be satisfied by an infinite number of molecular speed distributions that vary with location and direction. It can also be shown that to satisfy nonequilibrium momentum boundary conditions, energy may be transferred between the gas and surface during the interaction, with the magnitude and sign of the energy transfer varying with location on the cavity surface. Moreover, in a stationary state, the collective average energy of molecules leaving the entire cavity surface is equal to the collective average energy of molecules striking the entire cavity surface for a molecular flow cavity. Thus, by definition, the total energy of the gas contained in the cavity cannot change in a stationary state. However, the results (not included in this paper due to length) do indeed allow for local energy flow between the gas and any given surface element, e.g. energy may flow from $dA$ to the gas and from the gas to $dA'$. An actual stationary state solution will depend on additional properties of the cavity wall, e.g. its thickness, thermal conductivity, and temperature profile of the outer surface.

Thus, nonequilibrium stationary states that exhibit nonuniform flux densities and location-dependent energy transfer between the surface and gas can (and indeed, must) simultaneously satisfy flux balance, conservation of energy, and conservation of momentum. This result is intuitively expected (it is the physical basis for accommodation pumps [24]), and its demonstration lays the foundation for more intriguing subsequent results. To further set up the discussion of those results, we now consider an example.

## B. Example of hemispherical cavity

It is instructive to consider a specific example of a stationary state having a nonequilibrium nonuniform flux density solution. To quantitatively compute a flux-balancing solution, expressions for incoming and outgoing flux densities are needed. The outgoing flux density is a function of the outgoing angular flux distribution at the same location. The incoming flux density is a function of outgoing angular flux distributions elsewhere on the cavity surface, and the shape of the cavity. Thus, the specification of the cavity shape and the outgoing angular flux distribution function, when constrained by flux balance, uniquely determines the incoming flux density at every location on the cavity surface. The expressions necessary to compute a flux-balancing solution are derived in Appendix C for a cavity as shown in Fig. 2(a).

Because solutions depend on the cavity geometry, we consider the hemispherical cavity shown in Fig. 3(a) as an example. Since we have not yet considered the details of the gas-surface interaction physics, for the purposes of the present illustration we arbitrarily specify the form $x(r,\theta_r,\phi_r)$ of an outgoing angular flux distribution function, and compute the unique associated scaling function $y(r)$ that results in flux balance. Specifically, we let the outgoing angular flux distribution be of the form:

$$P_n(r,\theta_r,\phi_r) = F_{out}(r)\frac{(n+1)}{2\pi}\cos^n(\theta_r), \qquad (2)$$

where $n$, not constrained to be an integer, is selected to be the same for all surface elements of the hemispherical cavity. The normalization coefficient $(n+1)/2\pi$ is chosen so that, for any $n$, the outgoing flux density is $F_{out}(r)$ (see Eq. (C3)). The objective is, for a given value of $n$, to determine the function $F_{out}(r)$ in Eq. (2) that satisfies flux balance at all locations on the cavity surface.

To simplify notation, we let the location $r$ of a surface element of the cavity be parameterized by an angle $\sigma$ if $r$ is on the bounding hemisphere or a distance $u \equiv d/R$ if $r$ is on the bounding plane, as shown in Fig. 3(a). The resulting position-dependent flux density for the case $n = 3$ is derived in Appendix D, and is

$$F_{out-hemisphere}(\sigma) \approx 1.2926 - 0.3932\sigma^2 + 0.0393\sigma^4$$

$$F_{out-plane}(u) \approx 2.0745 - 2.0132u^2 + 0.3087u^4 \qquad (3)$$

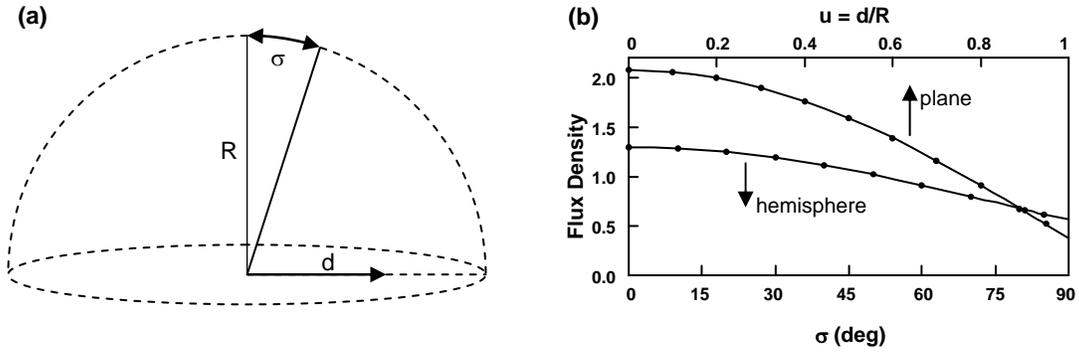

**FIG. 3.** (a) Flux-balancing solutions are found for a hemispherical cavity in the molecular flow regime. The location $r'$ of a surface element is parameterized by an angle $\sigma$ if it is on the bounding hemisphere or a distance $u = d/R$ if it is on the bounding plane. (b) Flux-balancing solution for $n = 3$ in Eq. (2). Solid lines indicate the outgoing flux density $F_{out}(r')$ resulting from Eq. (C3). Dots represent the incoming flux density $F_{in}(r')$ resulting from integrating Eq. (C2) over the cavity surface. Note that $F_{out}(r') = F_{in}(r')$, i.e. flux is balanced at each location.

We see from Fig. 3(b) that flux is indeed balanced. Specifically, the outgoing flux density, determined by evaluating Eq. (C3) using Eqs. (2) and (3), is plotted as solid lines. The incoming flux density, determined by evaluating Eq. (C2) using Eqs. (2) and (3), is plotted as dots. The incoming and outgoing flux densities are the same, and vary with location on the cavity surface. The ratio of the maximum flux density to the minimum flux density is approximately 5.

Further insight can be gained by plotting the position-dependent flux density as shown in Fig. 4. Specifically, the length of each line in Fig. 4 is proportional to the flux density plotted in Fig. 3(b). The flux density at the center of the bounding plane is the greatest, and at the edge of the bounding plane it is the least. The flux density also varies with position on the bounding hemisphere.  This flux density gradient, were it to physically occur in a nonequilibrium stationary state system, would have to be sustained by anisotropic gas-surface interactions.

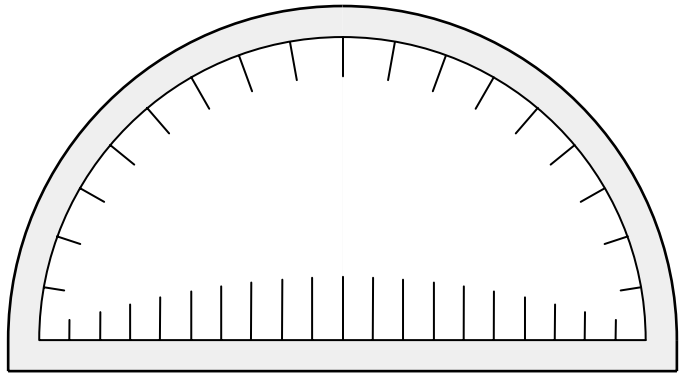

**FIG. 4.** The same data as in Fig. 3(b), but plotted more intuitively. The relative magnitude of the flux density $F(r')$ is indicated by the length of the lines. The flux density varies with location both on the bounding hemisphere and on the bounding plane for this stationary state solution.

Because the stationary state configuration depends, in part, on cavity geometry, the stationary state of a cavity will change in response to a change in cavity geometry. An important consequence of this property is now illustrated.  Consider two separate cavities, as shown in Fig. 5(a), each of which is in a stationary state.  When the cavities are connected by a port as in Fig. 5(b), the flux is initially not balanced at the port, and gas will flow from Cavity 1 to Cavity 2 until a new stationary state is achieved where the flux is balanced at the port as shown.  Finally, the flux configuration can be reversed by changing the location of the interconnecting port, as shown in Fig. 5(c).  Hence, the macroscopic configuration is completely reversible simply by opening and closing valves, i.e. by changing the geometric configuration of the cavity.

The analysis associated with Fig. 5 thus demonstrates that anisotropy at the microscopic scale (i.e. a non-cosine gas-surface interaction) can lead to reversibility at the macroscopic scale.

It is worthwhile to point out that the only value of $n$ in Eq. (2) that leads to a uniform flux density for the hemisphere is $n \equiv 1$, as expected.  As we saw earlier, the cosine distribution always gives rise to a position-independent flux density for any cavity geometry.

Clearly, non-cosine stationary state solutions such as that depicted in Figs. 4 and 5 must be nonequilibrium solutions, i.e. they are configurations where energy and momentum is exchanged between the gas and surface.  A conventional method to achieve the required anisotropic gas-surface interactions is to externally maintain a thermal gradient along the outer surface of the cavity.

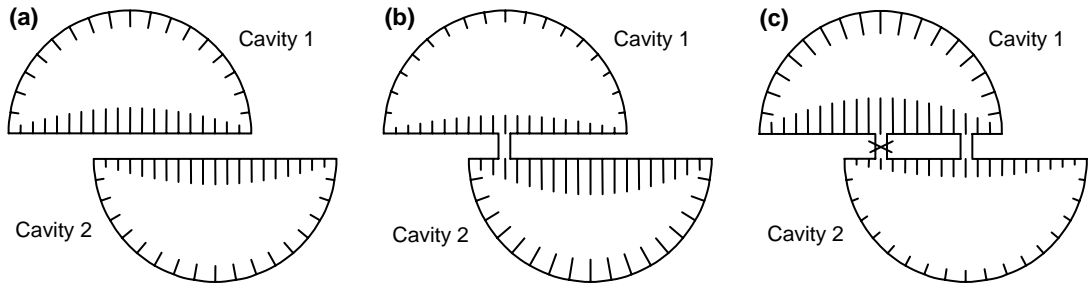

**FIG. 5.** Stationary state flux distributions can be geometry dependent.  (a) Two cavities are in identical stationary states.  When the cavities are connected by a port as in (b), gas flows from Cavity 1 to Cavity 2 until the flux is balanced through the port, resulting in a new stationary state.  When the first port is closed and a second port is opened (c), gas flows from Cavity 2 to Cavity 1 until flux is balanced through the second port, resulting in a different stationary state.

A simple experimental test for a non-cosine surface is illustrated by Fig. 6. Here, a suitable substrate such as a silicon wafer may be anisotropically etched to form a series of interconnected cavities shaped as hemi-cylinders. Connected to the array of cavities are cylindrical chambers covered with a membrane made of a deformable material such as a silicon nitride film. If the surfaces of the cavities are formed so they result in a non-cosine outgoing angular flux distribution, and the cavities are in the molecular flow regime, a nonuniform flux density will result (conceptually similar to Fig. 5). This will lead to a stationary state pressure difference between the chambers. A pressure difference may be detected by measuring the deflection of the membranes using an interferometer, for example. Many cavities may be connected in series to increase the pressure difference between the chambers to a value sufficient to cause a measurable membrane deflection. Note that there is no size constraint for the pressure-sensing chambers – for example, they could be a liter in volume and be in the viscous flow regime.

To assess the feasibility of the experiment suggested above, we note the following. Cavities with a maximum dimension of 5-10 μm (easily produced with modern micromachining technologies) will be in the molecular flow regime at a pressure of less than ~1 Torr. If the cavity surface provides an effective value of only $n = 1.02$ (see Eq. (2)), a string of 100 hemi-cylinder cavities connected in series will result in a flux ratio of greater than 2 between the input and output ports. A silicon nitride membrane 2 μm thick covering a round chamber 1.5 mm in diameter will deflect, in response to a pressure difference of only 0.1 Torr, approximately 300 nm; this is easily detectable with an interferometer. Thus, the system depicted in Fig. 6 can provide a very sensitive test of a stationary state non-cosine outgoing angular flux distribution. As indicated previously, a conventional method to achieve a non-cosine distribution is to maintain a thermal gradient on the cavity walls by an external heat source.

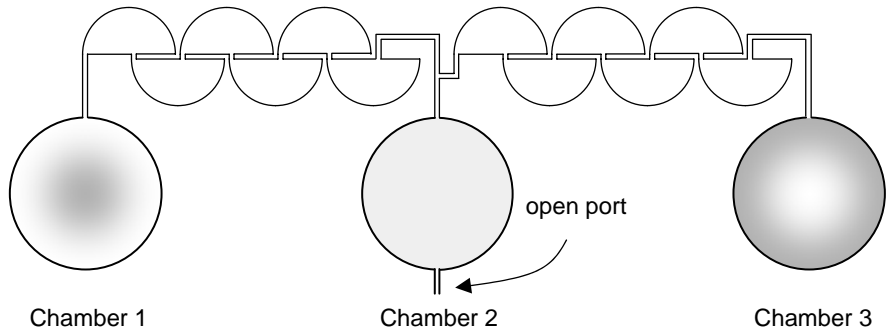

**FIG. 6.** Stationary state non-cosine outgoing angular flux distributions can be detected by observing deflections of membranes covering each of three chambers. The membrane covering Chamber 1 will deflect downward due to lower internal pressure (for $n > 1$), the membrane covering Chamber 2 will not deflect because the internal and external pressures are equal, and the membrane covering Chamber 3 will deflect upward due to higher internal pressure.

# V. DISCUSSION

The above analyses clearly demonstrate the existence of two distinct families of stationary states that satisfy flux balance, conservation of momentum, and conservation of energy. The first family is that of systems in conventional equilibrium, which are characterized by a cosine outgoing (and incoming) angular flux distribution. The second more intriguing family is that of nonequilibrium systems; these are characterized by non-cosine outgoing angular flux distributions. Of particular interest is what scenarios can lead to non-cosine outgoing angular flux distributions, thereby enabling macroscopically reversible systems.

## A. Self-sustaining nonequilibrium stationary state

The issue of how a nonequilibrium stationary state may be sustained is an extremely important one. Clearly, a nonequilibrium stationary state can be sustained for the cavity in Fig. 2(a) if a nonuniform temperature profile is maintained on the exterior surface using an external energy source, as is the case for an accommodation pump [24]. But now we consider whether a nonequilibrium configuration can result for the cavity shown in Fig. 7, where the exterior surface of the cavity wall is maintained at a uniform constant temperature, i.e. the cavity is immersed in an ideal isothermal bath. From our earlier analyses, we see that this question is equivalent to assessing whether or not a non-cosine outgoing angular flux distribution can occur at the inner surface as a stationary state with an isothermal external surface as a boundary condition.

There are numerous and well known experimentally-based reasons to anticipate that the nature of the inner surface could have a profound impact on the outgoing angular flux distribution. Nano-technological developments over the past decade have resulted in the routine creation of reduced-dimensionality surfaces [15,16], such as carbon nano-tubes [11] and quantum wires [14], that exhibit highly anisotropic optical [25], phonon [13,26,27], and electronic [28,29] properties. Quasi-equilibrium thermal desorption profiles can deviate quite significantly from a cosine form [30-32]. The thermodynamics and kinetics of reduced dimensionality surfaces can deviate from conventional flat or amorphous surfaces [12]. Various types of thermo molecular pumps require asymmetries in gas-surface interactions to function [24]. Even for optical processes, emission-absorption symmetry can be broken [33].

Thus, reduced-dimensionality surfaces known to exhibit various anisotropic properties are candidate surfaces that could possibly give rise to anisotropic (i.e. non-cosine) outgoing angular flux distributions. From our previous analyses, it is conceivable that such a non-cosine flux distribution could give rise to a nonuniform flux density and a thermal gradient on the interior surface, and could exist while the exterior surface is isothermal while at the same time satisfying flux balance, conservation of energy, and conservation of momentum. This would be a self-sustaining nonequilibrium stationary state if it physically occurred.

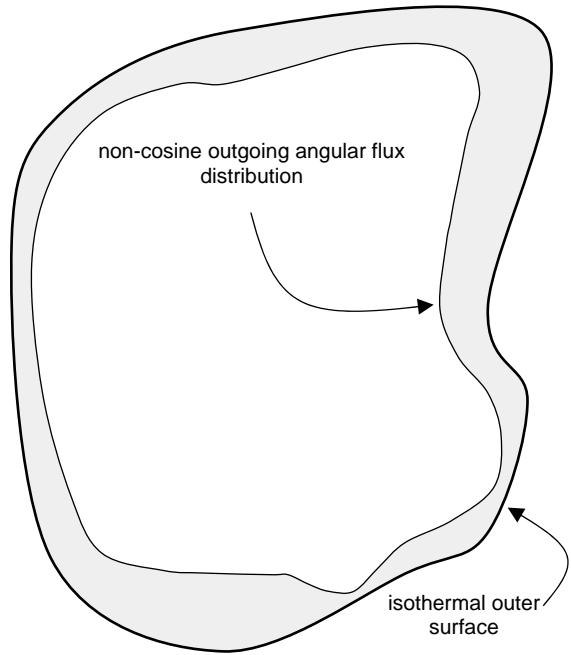

**FIG. 7.** A possibly self-sustaining nonequilibrium stationary state. The outer surface of the cavity wall is maintained as an isothermal surface, while a non-cosine outgoing angular flux distribution may occur on the inner surface of the cavity wall, possibly as a result of anisotropic interactions with a reduced-dimensionality surface. Such a system may exhibit continuous local energy redistribution, and may exhibit a non-isothermal inner surface.

A nonequilibrium stationary state self-sustained by anisotropic molecular interactions with a reduced-dimensionality surface would clearly allow the second law of thermodynamics to be circumvented (whether it would be "violated" or simply "circumvented" because of its inapplicability is a matter of semantics). Implementations such as those shown in Figs. 5 and 6 (and numerous variations) would permit reversible energy redistribution within a system immersed in an isothermal bath, and could permit sustained energy extraction from an isothermal bath on a macroscopic scale.

We are thus left with the following two possibilities for a cavity such as that shown in Fig. 7:
a) All possible inner surfaces must always exhibit precisely a cosine distribution, e.g. $n \equiv 1$ in Eq. (2). This results in a uniform flux distribution over the inner surface of the cavity, and the second law of thermodynamics always being satisfied.
b) Some surface can exist that may exhibit a non-cosine distribution. This would permit energy to be reversibly redistributed, "circumventing" the second law.

Obviously, the above two options are mutually exclusive. Because of the invulnerable position held by the second law, this leads us to a discussion of the cosine distribution.

### B. The cosine distribution

That all possible interior surfaces must exhibit a cosine outgoing angular flux distribution (for a cavity with an isothermal exterior surface) is a universal property that would seem to have a rather fundamental and obvious theoretical origin. That is not the case. There have indeed been extensive attempts to justify the cosine distribution over the past century. These continued attempts are a reflection of the tremendous skepticism with which scientists expect the cosine distribution to be the only one that results from complex gas-surface interaction physics and diverse boundary conditions. Indeed, Wenaas [19] clearly summarized the prevailing sentiment regarding the relationship between the cosine distribution and the second law, stating "While it may be possible to formulate some other argument demanding a cosine flux of scattered particles at equilibrium based on the second law, it would not be very satisfying to do so from the standpoint of understanding the details of the microscopic processes involved, …". Interestingly, all attempts known by the author to justify the cosine distribution have been unsuccessful apart from a starting point of the a-priori assumption of the second law. It is worthwhile to review several historical arguments, as summarized below.

Many authors have attempted to derive or justify the cosine distribution, and have often struggled to determine what comprises the most appropriate starting assumptions. To consider a few examples, Gaede [34] provided an argument why the cosine flux distribution was a required consequence of the second law. Clausing [35] later indicated the cosine distribution was a consequence of detailed balance (required by the second law). Wenaas [19] subsequently demonstrated that Gaede's proof was incorrect. In the same paper, Wenaas attempted to derive the cosine distribution from principles other than the second law. However, the very first step in his analysis was to assume that the molecular distributions are independent of location. This immediately constrains the distributions, solely from geometric arguments discussed earlier in the present paper, to be equilibrium distributions and to satisfy the second law. Furthermore, Wenaas stated "the assumption that the states of the surface are distributed randomly with respect to direction appears to be fundamental." It is not at all evident that assumption is applicable to reduced-dimensionality surfaces, whose properties were not known at the time of his work. Wenaas went on to show that reciprocity leads to the cosine law, and derived a reciprocity relationship based on a host of assumptions primarily associated with gas-surface interactions, including the above-mentioned fundamental constraints. Kuscer [17] demonstrated that detailed balance results if time reversal invariance is satisfied and the gas and wall are in thermal equilibrium. Like Wenaas, Kuscer also assumed the flux was position independent. Comsa [31,32] argued that various components of the outgoing distribution must deviate from cosine in form ("there is no a priori reason for the equilibrium angular distribution of desorbing molecules to be cosine" [32]) because the total distribution "must be cosine" [32], i.e. he simply adopted the prevailing view that the total distribution was cosine in form. Horodenski [18] provided an insightful assessment of detailed balance, questioning its universal applicability. Interestingly, even he steadfastly adhered to the prevailing perspective, and asserted, "the process of gas-surface scattering must be insensitive of geometry", referring to the geometry of the cavity. In the above-mentioned analyses, the assumptions are typically complex, all require a-priori constraints regarding the form of the gas distribution, and all are equivalent to simply assuming the second law.

The present paper provides the most direct second law-based derivation of the cosine distribution, and relies solely on the geometrical analysis of flux distributions. Specifically, isotropy and homogeneity of the molecular speed distribution require the flux distribution to be cosine in form, and to have the same magnitude at all surface locations. Alternatively, assuming the angular dependence of the outgoing angular flux distribution is independent of location (e.g. Eq. 2), a uniform flux density results only for the cosine distribution. The cosine distribution and the second law are clearly inextricably linked.

## C. Additional Insights

In lieu of the cosine distribution, one would anticipate that some fundamental principle (other than the second law) would exclude the possibility of a self-sustaining, surface induced, nonequilibrium stationary state. That appears to not be the case. For the sake of completeness, we now consider the essential topics of reversibility and causality.

*Reversibility*

The fundamental issue of concern, in regards to the second law of thermodynamics, pertains to reversibility; the second law is rooted in the observational requirement for irreversibility [36-41]. The chaotic hypothesis is a key requirement for maintaining irreversibility. The H-Theorem [5,6], which constrains the time-evolution of entropy, is derived from the Boltzmann equation. The validity of the Boltzmann equation depends on the chaotic hypothesis being satisfied, both for the gas and for the boundary conditions. "When proving that chaos is preserved in the limit, it is absolutely necessary to have a boundary condition compatible" with the same generalized form of the chaotic hypothesis that is assumed for the gas [5]. Boundary conditions not satisfying the chaotic hypothesis lead to a murkier situation. "There is no generally accepted definition of entropy in nonequilibrium stationary states." [42] In any case, it is clear that the statistics and sensitivity to initial conditions of chaotic systems prevents macroscopic reversibility at meaningful time scales. Only for small systems and short times has the second law been statistically "violated", as described by the Fluctuation Theorem (FT) and the Transient Fluctuation Theorem (TFT) [43,44] and measured in related experiments [45-47].

The method for achieving reversibility discussed in the present paper fundamentally circumvents the irreversibility limitations of conventional chaotic processes. The chaotic hypothesis cannot be a-priori assumed. A molecular flow cavity provides a way for microscopic anisotropy (non-cosine gas-surface interactions) to span to the macroscopic scale. Asymmetry in the cavity geometry may sustain nonequilibrium boundary conditions for a stationary state. Reduced-dimensionality surfaces may provide the required microscopic anisotropy.

We note the very important point that the second law is found to hold for all systems observed thus far. We also note that a system such as that depicted in Fig. 6, with the interior surfaces formed with a suitable reduced-dimensionality surface, has probably never been appropriately observed in the molecular flow regime while immersed in an isothermal bath.

*Causality*

An even more fundamental issue than reversibility is that of causality. Consider again the now-familiar molecular flow cavity depicted in Fig. 7, where the exterior is maintained as an isothermal surface. For initial conditions, let the inner surface be isothermal and let the gas be isotropic and homogeneous and at the same temperature as the inner surface. After a period of time, a stationary state is achieved. What physically "causes" the stationary state outgoing angular flux distribution to be what it is (even if it is cosine)? The "cause" must be the physics of the gas-surface interactions, coupled with the boundary conditions for those interactions. That the outgoing angular flux distribution for the ensuing stationary state must be uniformly cosine, for all possible surfaces and cavity geometries, remains a truly fundamental and unresolved mystery.

**D. Two foundational opportunities**

To conclude the discussion, we review several key facts. Flux balance, conservation of momentum, and conservation of energy do not constrain the outgoing distribution to be cosine. An infinite number of non-cosine solutions exist, and they depend on the geometry of the cavity. Reduced-dimensionality surfaces exist that experimentally exhibit numerous highly anisotropic properties. There is no existing justification for the cosine distribution other than the second law itself. This leads us to two foundational opportunities for the motivated reader:

*Opportunity #1*
A fundamental derivation for why the outgoing angular flux distribution must be cosine is desperately needed – one that is not a cleverly obscured circular argument based on a-priori assumptions equivalent to requiring the second law to be satisfied. One might anticipate such a fundamental and universal result, applicable even to highly anisotropic reduced-dimensionality surfaces, would be profoundly simple to derive. It has thus far eluded discovery.

*Opportunity #2*
Build a system with a non-cosine surface, and experimentally demonstrate the first truly macroscopic sustained circumvention of the second law of thermodynamics. Such a circumvention of the second law would result from microscopic anisotropy that leads to macroscopic reversibility.

Success at either of these mutually exclusive opportunities will profoundly impact our understanding of thermodynamic principles.

## VI. SUMMARY

We quantitatively investigated thermodynamic implications of anisotropic gas-surface interactions in a molecular flow cavity. In addition, we provided a detailed and insightful analysis of the question "must all surfaces, including reduced-dimensionality surfaces, be inherently randomizing?" Several startling results clearly motivate continuing and serious research into the fundamental behavior of gases in the molecular flow regime.

It was demonstrated that anisotropy at the microscopic scale can lead to reversibility at the macroscopic scale. Specifically, a non-cosine outgoing angular flux distribution results in a stationary state that exhibits a nonuniform flux density that depends on the shape of the cavity. Such a geometry-dependent nonuniform flux density may give rise to a macroscopically reversible stationary state such as that illustrated by Fig. 5.

It was shown that existing derivations of the cosine distribution require assumptions equivalent to assuming the second law. No derivations were found by the author that were based on fundamental mechanisms independent of the second law.

It was shown that, independent of the second law itself, there is no reason to exclude the possibility of a self-sustaining nonequilibrium stationary state for a system immersed in an isothermal bath. It was mathematically demonstrated that such a system may simultaneously satisfy flux balance, conservation of momentum, and conservation of energy. A candidate experiment was suggested to demonstrate such a system, as illustrated by Fig. 6.

It was proposed that reduced-dimensionality surfaces, known to exhibit a multitude of anisotropic properties, are candidate surfaces to produce non-cosine outgoing angular flux distributions necessary for a self-sustaining nonequilibrium stationary state system.

## ACKNOWLEDGMENTS

The author is grateful to colleagues who provided helpful comments and encouragement regarding this work. Special thanks to W. M. Miller and S. Montague.

# APPENDIX A – Isotropy and Homogeneity Require the Cosine Flux Distribution

## A. Definitions

In this and all following analyses, we do not attempt to model molecular flux at the level of individual molecules. Instead, we address it for an ensemble of molecules interacting with a surface element. In particular, we consider a time interval long enough to sample the average molecular flux associated with a surface element in a stationary state. The same is done for all surface elements of a cavity. Such averaging is appropriate for a stationary state system.

We begin with several important definitions regarding terminology and coordinate systems that will be used for this and other analyses. Consider a surface element *dA'* in Fig. 2(a), located at position *r'*, and forming part of a boundary of a closed cavity containing a gas in the molecular flow regime. Over a unit time interval, a certain number of molecules strike *dA'* from all directions; this is the incoming molecular flux to *dA'*. One of the surface elements contributing to this incoming flux is *dA* located at position *r*, which is a distance $\rho$ from *dA'*. The straight line connecting the two surface elements forms an angle $\theta_{r'} = \theta_{dA' \to dA} = \beta$ relative to the normal of *dA'* at *r'*, and an angle $\theta_r = \theta_{dA \to dA'} = \alpha$ relative to the normal of *dA* at *r*. Similarly, the line connecting the two surface elements form azimuthal angles $\phi_{r'} = \phi_{dA' \to dA}$ at *r'* and $\phi_r = \phi_{dA \to dA'}$ at *r*. For surface element *dA* located at *r*, the outgoing angular flux distribution function, i.e. the outgoing flux per unit area (i.e. flux density) per unit solid angle in direction ($\theta_r$, $\phi_r$) is designated as $P(r, \theta_r, \phi_r)$.

The solid angle from *dA* subtended by surface element *dA'* is

$$d\Omega_{A \to A'} = \frac{dA' \cos(\theta_{dA' \to dA})}{\rho^2}. \tag{A1}$$

The solid angle from *dA'* subtended by surface element *dA* is

$$d\Omega_{A' \to A} = \frac{dA \cos(\theta_{dA \to dA'})}{\rho^2}. \tag{A2}$$

The flux impinging on *dA'* from surface element *dA* is

$$\delta Flux_{A \to A'} = P(r, \theta_{dA \to dA'}, \phi_{dA \to dA'}) \, dA \, d\Omega_{A \to A'} = P(r, \theta_{dA \to dA'}, \phi_{dA \to dA'}) \, dA \, \frac{dA' \cos(\theta_{dA' \to dA})}{\rho^2}. \tag{A3}$$

The flux impinging on *dA* from surface element *dA'* is

$$\delta Flux_{A' \to A} = P(r',\theta_{dA' \to dA},\phi_{dA' \to dA})\, dA'\, d\Omega_{A' \to A} = P(r',\theta_{dA' \to dA},\phi_{dA' \to dA})\, dA'\, \frac{dA \cos(\theta_{dA \to dA'})}{\rho^2}.$$

(A4)

**B. Implications of isotropy and homogeneity**

Molecular speed isotropy requires the flux to be balanced between surface elements *dA* and *dA'*. This requires the equality of Eqs. (A3) and (A4), i.e. $\delta Flux_{A \to A'} = \delta Flux_{A' \to A}$, leading to

$$\frac{P(r,\theta_{dA \to dA'},\phi_{dA \to dA'})}{\cos(\theta_{dA \to dA'})} = \frac{P(r',\theta_{dA' \to dA},\phi_{dA' \to dA})}{\cos(\theta_{dA' \to dA})} = K.$$

(A5)

Since the first and second terms are functions of different independent quantities, each term must be a constant. The assumption of isotropy and homogeneity requires the constant *K* to be the same for each and every pair of surface elements in the cavity. Thus, the outgoing angular flux distribution at each and every location *r* must be:

$$P(r,\theta_r,\phi_r) = K \cos(\theta_r).$$

(A6)

The angular flux distribution is independent of location on the cavity surface, and has no azimuthal dependence. This result is also independent of how one chooses to define the shape of the surface at a microscopic level, i.e. it is independent of the nanostructure, even at the molecular level, provided one can perform the averaging necessary to define a distribution. This is solely a geometrical result applicable to any molecular flow cavity with an isotropic and homogeneous molecular speed distribution, i.e. where flux is balanced between any pair of surface elements. Because the flux is balanced between any two surface elements, Eq. (A6) is also the incoming angular flux distribution.

# APPENDIX B – Flux-Balancing Solutions Exist That Exhibit Nonuniform Flux Density

In Appendix A we considered the consequence of requiring flux balance between individual pairs of surface elements, as results from isotropy and homogeneity. In contrast, the objective here is to assess the implications of flux balance at each individual surface element. One might anticipate that the family of allowable solutions would be extremely constrained. That is not the case. We discover this by replacing the inner surface of the curved cavity of Fig. 2(a) with a cavity bounded by a finite number of facets $m$, as shown in Fig. B-1. For each facet, there are $m$-1 directions that point to the remaining $m$-1 facets. Hence, there are $m$-1 outgoing flux components for each facet. Since there are $m$ facets, there are a total of $m(m$-$1)$ outgoing flux components for the entire cavity. Flux balance requires that for each of the $m$ facets, the sum of the outgoing flux components equals the sum of the incoming flux components. This leads to a total of $m$ simultaneous equations containing $m(m$-$1)$ parameters (flux components). Thus, $m^2$-$2m$ flux components may be arbitrarily chosen, with the remaining $m$ flux components being selected to satisfy the $m$ flux balance equations. The existence of $m^2$-$2m$ excess parameters permits extreme flexibility in choosing a flux-balancing solution, particularly when the number of facets $m$ is asymptotically large. Consequently, flux balance can be satisfied without requiring the flux density to be the same at all locations, and without requiring any particular form for the outgoing angular flux distribution.

To consider an important example, let the outgoing angular flux distribution be a different arbitrary function for each of the $m$ facets. Flux balance may be achieved simply by appropriately scaling the magnitude of the distribution function for each facet. The *m* scaling factors, one for each facet, provides precisely the number of degrees of freedom to uniquely satisfy the *m* flux balance equations. Stated explicitly, simple position-dependent scaling is all that is necessary to satisfy flux balance for any form of arbitrary position-dependent outgoing angular flux distribution.

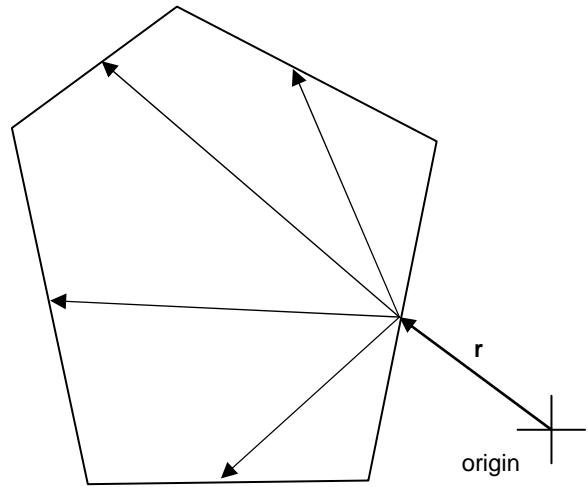

**FIG. B-1.** This cavity is bounded by $m = 5$ facets. For each facet, there are $m – 1$ (=4) outgoing flux components, one for each receiving facet. For the entire cavity, there are a total of $m(m – 1)$ (=20) outgoing flux components. The flux components are constrained by the requirement of flux balance at each facet, i.e. the total incoming flux must equal the total outgoing flux in a stationary state. In general, the number of facets $m$ may be asymptotically large.

In the continuous limit, i.e. as $m \to \infty$, the above example leads to the following. Let $x(\mathbf{r}, \theta_r, \phi_r)$ be an arbitrary function of position and angle. Let $y(\mathbf{r})$ be a position-dependent scaling function. The outgoing angular flux distribution function $P(\mathbf{r}, \theta_r, \phi_r) \equiv y(\mathbf{r}) x(\mathbf{r}, \theta_r, \phi_r)$ satisfies flux balance for a unique function $y(\mathbf{r})$. Thus, for each function $x(\mathbf{r}, \theta_r, \phi_r)$, one and only one corresponding scaling function $y(\mathbf{r})$ results in flux balance for all locations on a cavity surface.

The above result is of fundamental significance. It is anticipated that the detailed physics and boundary conditions of gas-surface interactions will determine the underlying *form* $x(\mathbf{r}, \theta_r, \phi_r)$ of the outgoing angular flux distribution function. *Any* such form satisfies flux balance at each surface element simply by position-dependent multiplicative scaling, i.e. simply by the existence of a gradient in the flux density.

# APPENDIX C – Method to Find a Flux-Balancing Solution for an Arbitrarily Shaped Cavity

To avoid being distracted by notational details, we drop the explicit reference to the azimuthal angle $\phi$, which is equivalent to assuming azimuthal symmetry for the outgoing angular flux distribution at each location on the surface. This simplification (assumption) will not impact our fundamental conclusions.

Referring to Fig. 2, the total incoming flux to surface element $dA'$ is given by integrating the outgoing contributions from all line-of-sight surface elements $dA$ forming the surface of the cavity. Specifically, the flux striking $dA'$ at $r'$ is given by integrating Eq. (A3) over all line-of-sight surface elements $dA$ of the cavity:

$$Flux(r') = \int_A \delta\, Flux_{A \to A'} = \iint_A P(r,\alpha) \frac{dA'\cos(\beta)}{\rho^2}\, dA. \tag{C1}$$

Note that we have simplified notation by defining $\alpha = \theta_{dA \to dA'}$ and $\beta = \theta_{dA' \to dA}$. The total flux per unit area impinging on surface element $dA'$, i.e. the incoming flux density, is

$$F_{in}(r') = Flux(r')/dA' = \iint_A P(r,\alpha) \frac{\cos(\beta)}{\rho^2}\, dA. \tag{C2}$$

Note that evaluation of Eq. (C2) requires specification of the cavity geometry. The integration is performed only over the portion of the cavity with line-of-sight flux contributions to the location $r'$.

Over a unit time interval, a certain number of molecules leaves surface element $dA'$; this is the outgoing flux. The corresponding outgoing flux per unit area, or outgoing flux density, is computed by integrating, over all solid angles $\Omega$ above the surface, the outgoing angular flux distribution function $P(r',\beta)$:

$$F_{out}(r') \equiv \iint_\Omega P(r',\beta)\, d\Omega. \tag{C3}$$

Flux balance only requires the equality of Eqs. (C2) and (C3) for every surface location $r'$ on the cavity. Note that flux balance does *not* require that the flux density be the *same* at every surface location.

It is useful to note that a particular benefit of the above formulation is that it does not involve the molecular speed distribution. This can be seen by the analogy of a pitcher throwing balls to a catcher. In a stationary state, the number of balls received by the catcher over some time interval depends only on the frequency at which balls are thrown by the pitcher, and not the speed of the balls thrown.

# APPENDIX D – Flux-Balancing Solution for a Hemispherical Cavity

We consider a family of non-cosine flux-balancing solutions of the form given by Eq. (2). In general, the flux density $F_{out}(\mathbf{r'})$ may vary with position on the cavity surface. Our objective is now to find a function $F_{out}(\mathbf{r'})$ that results in $F_{in}(\mathbf{r'}) = F_{out}(\mathbf{r'})$ at all points on the cavity. From our earlier analysis regarding general solutions, we know that, for each value of $n$, a unique solution $F_{out}(\mathbf{r'})$ does indeed exist that satisfies flux balance.

Flux-balancing solutions for the flux density depend on the geometry of the cavity. For the present example, we select a hemispherical cavity of radius $R$ as shown in Fig. 3(a). The location of a surface element whose flux balance is under consideration is parameterized by an angle $\sigma$ if $\mathbf{r'}$ is on the bounding hemisphere or a distance $u \equiv d/R$ if $\mathbf{r'}$ is on the bounding plane. A solution for a given $n$ can be obtained by expressing $F_{out}(\mathbf{r'})$ as a power series, and keeping as many terms as necessary for the precision desired.

## A. Point on bounding hemisphere

We first consider the incoming flux density at an arbitrary location on the bounding hemisphere of the cavity. Note that flux is received from both the hemisphere, and the planar boundary. We will compute the two contributions separately, and add the results to get the total incoming flux density as a function of location $\sigma$.

The incoming flux density from the hemisphere can be computed using the coordinate system and definitions depicted in Fig. D-1. Equation (C2) becomes, using Eq. (2)

$$F_{in-from\,hemi}(r') = \int_{\phi=0}^{2\pi} \int_{\theta=0}^{\pi/2} F_{out}(\theta) \frac{(n+1)}{2\pi} \cos^n(\alpha) \frac{\cos(\beta)}{\rho^2} R^2 \sin(\theta)\, d\theta\, d\phi. \tag{D1}$$

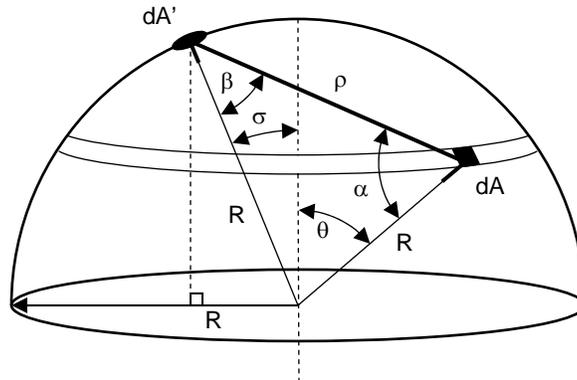

**FIG. D-1.** Geometric definitions used for computing flux-balaincing solutions for a hemispherical cavity. Incoming flux contributions to a point on the bounding hemisphere originating from the bounding hemisphere are considered.

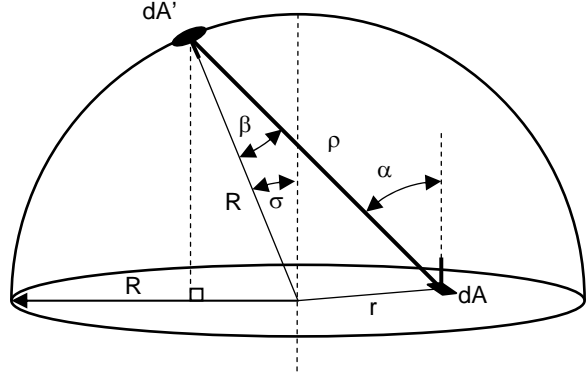

**FIG. D-2.** Incoming flux contributions to a point on the bounding hemisphere originating from the bounding plane are considered.

Using $\alpha = \beta$, $\rho = 2R\cos(\alpha)$, and $(\rho/R)^2 = 2[1 - \cos(\sigma)\cos(\theta) - \sin(\sigma)\sin(\theta)\cos(\phi)]$, we obtain

$$F_{in-from\,hemi}(\sigma) = \int_{\phi=0}^{2\pi}\int_{\theta=0}^{\pi/2} F_{out}(\theta)\frac{(n+1)}{8\pi}\left[\frac{1-\cos(\sigma)\cos(\theta)-\sin(\sigma)\sin(\theta)\cos(\phi)}{2}\right]^{(n-1)/2}\sin(\theta)\,d\theta\,d\phi$$

(D2)

The incoming flux density from the planar boundary can be computed using the geometrical definitions as shown in Fig. D-2. Equation (C2) becomes, using Eq. (2),

$$F_{in-from\,plane}(r') = \int_{\phi=0}^{2\pi}\int_{r=0}^{R} F_{out}(r)\frac{(n+1)}{2\pi}\cos^n(\alpha)\frac{\cos(\beta)}{\rho^2}r\,dr\,d\phi. \tag{D3}$$

Observing that $\rho\cos(\alpha) = R\cos(\sigma)$, $(\rho/R)^2 = 1 + (r/R)^2 - 2(r/R)\sin(\sigma)\cos(\phi)$, and $(r/R)^2 = 1 + (\rho/R)^2 - 2(\rho/R)\cos(\beta)$, we obtain

$$F_{in-from\,plane}(\sigma) = \int_{\phi=0}^{2\pi}\int_{u=0}^{1} F_{out}(u)\frac{(n+1)}{2\pi}\cos^n(\sigma)\frac{[1-u\sin(\sigma)\cos(\phi)]}{[1+u^2-2u\sin(\sigma)\cos(\phi)]^{(n+3)/2}}u\,du\,d\phi.$$

(D4)

where $u = r/R$.

The total incoming flux density impinging on the surface element *dA'* on the bounding hemisphere is given by summing the contribution from the hemisphere and from the plane, namely by adding Eqs. (D2) and (D4):

$$F_{in-to\,hemisphere}(\sigma) = F_{in-from\,hemi}(\sigma) + F_{in-from\,plane}(\sigma). \tag{D5}$$

## B. Point on bounding plane

We now consider the incoming flux density at an arbitrary location on the bounding plane of the cavity shown in Fig. D-3. As with any point on the bounding plane, the only contribution to the incoming flux density is from the bounding hemisphere. No flux from the plane containing $dA'$ impinges on $dA'$. Thus we have, from Eqs. (C2) and (2),

$$F_{in-to\,plane}(r') = \int_{\phi=0}^{2\pi} \int_{\theta=0}^{\pi/2} F_{out}(\theta) \frac{(n+1)}{2\pi} \cos^n(\alpha) \frac{\cos(\beta)}{\rho^2} R^2 \sin(\theta)\, d\theta\, d\phi. \quad (D6)$$

Observing that $\rho \cos(\beta) = R \cos(\theta)$, $(\rho/R)^2 = 1 + (d/R)^2 - 2(d/R)\sin(\theta)\cos(\phi)$, and $\cos(\alpha) = [1 - (d/R)\sin(\theta)\cos(\phi)]/(\rho/R)$, we obtain, with $u = d/R$,

$$F_{in-to\,plane}(u) = \int_{\phi=0}^{2\pi} \int_{\theta=0}^{\pi/2} F_{out}(\theta) \frac{(n+1)}{2\pi} \frac{[1 - u\,\sin(\theta)\cos(\phi)]^n \sin(\theta)\cos(\theta)\, d\theta\, d\phi}{[1 + u^2 - 2u\,\sin(\theta)\cos(\phi)]^{\frac{n+3}{2}}}. \quad (D7)$$

## C. Numerical solution

To evaluate the incoming fluxes to the hemisphere (Eq. (D5)) and plane (Eq. (D7)), one must select outgoing flux densities $F_{out}(\sigma)$ and $F_{out}(u)$, where $\sigma$ and $u = d/R$ indicate locations on the cavity surface. We approximate these functions as power series

$$F_{out}(\sigma) \approx a_0 + a_1\sigma + a_2\sigma^2 + a_3\sigma^3 + ... \quad (D8)$$

$$F_{out}(u) \approx b_0 + b_1 u + b_2 u^2 + b_3 u^3 + ... \quad (D9)$$

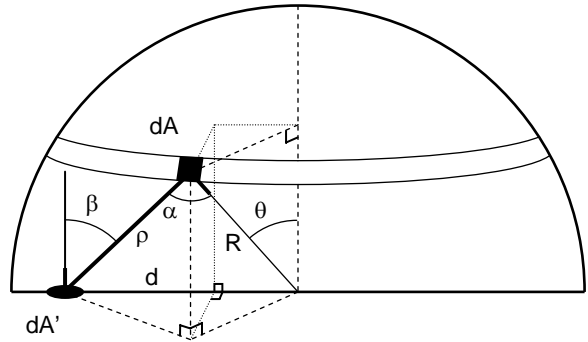

**FIG. D-3.** Incoming flux contributions to a point on the bounding plane originating from the bounding hemisphere are considered.

Flux balance requires the incoming flux to equal the outgoing flux at every surface location, i.e. Eqs. (D5) and (D8) must equal on the hemisphere, and Eqs. (D7) and (D9) must equal on the plane. The coefficients of Eqs. (D8) and (D9) are selected to satisfy flux balance.

The flux-balancing solution for $n = 3$ is well-approximated, for the hemisphere and plane respectively, by:

$$F_{out-hemisphere}(\sigma) \approx 1.2926 - 0.3932\sigma^2 + 0.0393\sigma^4 \tag{D10}$$

and

$$F_{out-plane}(u) \approx 2.0745 - 2.0132u^2 + 0.3087u^4. \tag{D11}$$

The results, shown in Fig. 3(b), were obtained by requiring flux balance at five locations: $\sigma = 0°$ and $65°$, and $u = 0$, 0.4, and 0.8. We also applied the constraint that the total flux impinging on the cavity is $3\pi$, which is the same as for a uniform flux distribution of unit magnitude. This last constraint is not a fundamental requirement, but is a convenient normalization method.

The few terms in Eqs. (D10) and (D11) are sufficient to provide the flux balance accuracy quantitatively shown in Fig. 3(b). Different values of $n$ may require different numbers of terms, depending on the accuracy desired.

# REFERENCES

1. M. Kinslow, AIAA Journal **14**,1358 (1976).
2. G. Ertl, Pure and Appl. Chem. **61**, 1001 (1989).
3. C. Cerignani and M. Lampis, AIAA Journal **35**, 1000 (1997).
4. C. Stampfi, H. Kreuzer, S. Payne, H. Pfnur, and M. Scheffler, Phys. Rev. Lett. **83**, 2993 (1999).
5. Carlo Cercignani, *Rarefied Gas Dynamics* (Cambrige University Press, 2000).
6. Hans Christian Ottinger, *Beyond Equilibrium Thermodynamics* (John Wiley & Sons, 2005).
7. C. E. Siewert and F. Sharipov, Physics of Fluids **14**, 4123 (2002).
8. G. Fan and J. Manson, Phys. Rev. B **72**, 085413 (2005).
9. A. Horodenski, J. Vac. Sci. Technol. A **3**, 39 (1985).
10. D. P. Sheehan, Phys. Rev. E **57**, 6660 (1998).
11. S. Iijima, Nature **354**, 56 (1991).
12. W. Widdra, P. Trischberger, W. Frieb, D. Menzel, S. Payne, and H. Kreuzer, Phys. Rev. B **57**, 4111 (1998).
13. M. Dresselhaus and P. Eklund, Adv. Phys. **49**, 705 (2000).
14. H. Yeom, J. of Electron Spectroscopy and Related Phenomena **114**, 283 (2001).
15. E. Plummer, Ismail, R. Matzdorf, A. Melechko, J. Pierce, and J. Zhang, Surf. Sci. **500**, 1 (2002).
16. L. Hueso and N. Mathur, Nature **427**, 301 (2004).
17. I. Kuscer, Surf. Sci. **25**, 225 (1971).
18. A. Horodenski, Phys. Lett. A **122**, 295 (1987).
19. E. Wenaas, J. Chem. Phys. **54**, 376 (1971).
20. J. Rowlinson, Mol. Phys. **103**, 2821 (2005).
21. E. Dahlberg, J. Phys. A **6**, 1800 (1973).
22. J. Blatt and A. Opie, J. Phys. A **7**, L113 (1974).
23. M. Knudsen, Ann. Phys. **48**, 1113 (1915).
24. J. Hobson and D. Salzman, J. Vac. Sci. Technol. A **18**, 1758 (2000).
25. Y. Murakami, S. Chiashi, Y. Miyauchi, M. Hu, M. Ogura, T. Okubo, and S. Maruyama, Chem. Phys. Lett. **385**, 298 (2004).

26. J. Hone, B. Batlogg, Z. Benes, A. Johnson, and J. Fischer, Science **289**, 1730 (2000).
27. J. Sauvajol, E. Anglaret, S. Rols, and L. Alvarez, Carbon **40**, 1697 (2002).
28. J. Jiang, J. Dong, and D. Xing, Phys. Rev. B **65**, 245418 (2002).
29. L. Alvarez, A. Righi, T. Guillard, S. Rols, E. Anglaret, D. Laplaz, and J. Sauvajol, Chem. Phys. Lett. **316**, 186 (2000).
30. U. Leuthausser, Phys. Rev. B **36**, 4672 (1987).
31. G. Comsa, J. Chem. Phys. **48**, 3235 (1968).
32. G. Comsa and R. David, Surf. Sci. Rep. **5**, 145 (1985).
33. M. Scully, Phys. Rev. Lett. **87**, 220601 (2001).
34. W. Gaede, Ann. Phys. **41**, 289 (1913).
35. P. Clausing, Ann. Phys. **4**, 533 (1930).
36. I. Prigogine, Physica A **263**, 528 (1999).
37. G. Gallavotti, J. Stat. Phys. **78**, 1571 (1994).
38. J. Lebowitz, Physica A **263**, 516 (1999).
39. J. Lebowitz, Rev. Mod. Phys. **71**, S346 (1999).
40. J. Gemmer, A. Otte, and G. Mahler, Phys. Rev. Lett. **86**, 1927 (2001).
41. A. Nikulov and D. Sheehan, Entropy **6**, 1 (2004).
42. G. Gallavotti and E. Cohen, Phys. Rev. E **69**, 035104 (2004).
43. J. Evans, E. Cohen, and G. Morriss, Phys. Rev. Lett. **71**, 2401 (1993).
44. E. Mittag and D. Evans, Phys. Rev. E **67**, 026113 (2003).
45. G. Wang, E. Sevick, E. Mittag, D. Searles, and D. Evans, Phys. Rev. Lett. **89**, 050601 (2002).
46. D. Carberry, J. Reid, G. Wang, E. Sevick, D. Searles, and D. Evans, Phys. Rev. Lett. **92**, 140601 (2004).
47. C. Bustamante, J. Liphardt, and F. Ritort, Physics Today **58**, 43 (2005).